\documentclass[pra,10pt]{revtex4}
\usepackage{color}
\usepackage{psfrag} 
\usepackage{graphicx}
\usepackage{amsmath}
\usepackage{amssymb}

\begin{document}

{\color{blue}

\title{Zeta Function Regularization of Infrared Divergences in
Bose-Einstein Condensation} 
\author{Adriaan M. J. Schakel}
\email{schakel@boojum.hut.fi}
\affiliation{Institut f\"ur Theoretische Physik \\ Freie Universit\"at
Berlin \\ Arnimallee 14, 14195 Berlin, Germany}

\begin{abstract}
The perturbative calculation of the effect of fluctuations in the
\textit{nonzero} frequency modes of a weakly interacting Bose gas on the
condensation temperature is reviewed.  These dynamic modes, discarded in
most of the recent studies, have a temperature-induced energy gap that
allows for a perturbative approach.  The simple, yet powerful algorithm used
to calculate the effect in a high-temperature expansion in conjunction with
zeta function regularization of infrared divergences is explained in detail.
The algorithm is shown to be reliable by demonstrating that it reproduces
known results for a series of examples.  With two-loop contributions
properly included, the dynamic modes are seen not to lead to a shift in the
condensation temperature, thus revising our earlier finding obtained at one
loop.
\end{abstract}
  
\date{\today}

\maketitle

\section{Introduction}
The past five years or so have seen numerous studies aimed to determine
the condensation temperature of a weakly interacting Bose gas (for
references and critical discussions of the various approaches, see
Ref.~\cite{Baymetal_rev}).  These studies, driven by experiments on
Bose-Einstein condensation (BEC), were based on a variety of analytic,
computational, and experimental methods.  With only a few exceptions,
however, the theoretical studies took the same starting point of
considering only the static part of the quantum field describing the
bosons.  The rationale is that the dynamic modes have a
temperature-induced energy gap, making them irrelevant compared to the
static mode, for which that gap vanishes, when studying the
long-distance critical behavior of the system close to the phase
transition.  With the gapped thermal modes decoupled, the microscopic
quantum theory reduces to a static, classical theory, assumed to
properly describe this thermal phase transition in equilibrium.
Universal properties, such as the critical exponents can indeed be
computed from this naive theory, as is well known.

Critical temperatures are, however, not universal.  Although phase
transitions occurring in different systems may display the same
long-distance behavior characterized by the same values of the critical
exponents, they in general take place at different temperatures.  For
example, BEC in a \textit{relativistic} free Bose gas is in the same
universality class as BEC in its nonrelativistic counterpart, yet the
condensation temperatures differ in both systems \cite{SP}.  This difference
arises because the critical temperatures of these noninteracting systems are
determined entirely by the dynamic modes (see Sec.~\ref{sec:highT}) whose
spectrum differ for the relativistic and nonrelativistic Bose gases.  The
universal properties, on the other hand, are determined by the static modes,
which are similar for both gases.  Whereas universal quantities are largely
independent of the dynamics and details of the underlying theory,
nonuniversal quantities in principle depend on them.  The critical
temperatures of lattice models are even sensitive to such short-distance
details as the precise lattice used.

The proper way to proceed when calculating nonuniversal properties is well
established.  One has to go down one level and start from the microscopic
model describing the quantum system under consideration and perturbatively
integrate out the irrelevant degrees of freedom (the dynamic modes when
considering BEC).  One thus obtains an effective theory in terms of the
remaining, relevant degree of freedom (the static mode in case of BEC), with
the coefficients being determined by the irrelevant degrees of freedom that
have been integrated out.  This \textit{effective field program} was carried
out for the problem at hand almost a decade ago \cite{effbos}, with the
conclusion that the dynamic modes lead to a shift in the condensation
temperature of a weakly interacting Bose gas.  The purpose of this paper is
to revise that finding.  Namely \cite{Arnold}, the shift obtained at one
loop is canceled by a two-loop contribution not considered in
Ref.~\cite{effbos}.  There, only the leading contribution in a
high-temperature expansion of the two-loop Feynman diagrams was considered,
whereas the canceling terms appear in the next order.  To show that it is
this omission and not, as has recently been suggested \cite{Baymetal_rev},
the method used that led to the incorrect conclusion, we clarify and justify
in detail the perturbative approach of Ref.~\cite{effbos}.  To appreciate
the approach, we first recall some essentials of finite-temperature field
theory.

\subsection{Finite Temperature}
The properties of a quantum system at finite temperature can be studied
\cite{FW} by going over to imaginary time $t \rightarrow - \mathrm{i}
\tau$, with $\tau$ restricted to the interval $0 \leq \tau \leq \hbar
\beta$, where $\beta=1/k_\mathrm{B} T$ is the inverse temperature.  The
time dimension thus becomes compactified and, consequently, the energy
variable $\hbar k_0$ discrete $k_0 \to \mathrm{i} \omega_n$, with
$\omega_n$ the so-called Matsubara frequencies, 
\begin{equation} \label{bcs:Mats}
\hbar \omega_n = \left\{ \begin{array}{ll}
2 \pi \beta^{-1} n       & {\rm for \;\; bosons}    \\
\pi \beta^{-1} (2n + 1)  & {\rm for \;\; fermions},
\end{array}        \right.                                       
\end{equation} 
where $n$ is a (positive or negative) integer.  The difference between
the bosonic and fermionic frequencies arises because a bosonic field
$\phi(\tau, \mathbf{x})$ satisfies periodic boundary conditions
$\phi(\hbar \beta,\mathbf{x})= \phi(0,\mathbf{x})$, whereas a fermionic
field satisfies antiperiodic ones $\psi(\hbar \beta, \mathbf{x})=
-\psi(0,\mathbf{x})$.  Being (anti)periodic, the fields can be expanded
in a Fourier series as (with a similar equation for $\psi$):
\begin{equation} 
\label{fs}
\phi (\tau, {\bf x}) = \frac{1}{\hbar \beta} \sum_{n=-\infty}^\infty \int
\frac{\mathrm{d}^d k}{(2\pi)^d} \, \mathrm{e}^{- \mathrm{i} \omega_n \tau +
\mathrm{i} \mathbf{k} \cdot \mathbf{x}} \, \phi_n (\mathbf{k}),
\end{equation} 
where $\phi_n (\mathbf{k}) := \phi (\omega_n, \mathbf{k})$.

An integral over the energy variable $\hbar k_0$ at the absolute zero of
temperature becomes a sum over the Matsubara frequencies at finite
temperature:
\begin{equation}
\int \frac{\mathrm{d} k_0}{2\pi} \, g(k_0) \to \frac{\mathrm{i}}{\hbar
\beta} \sum_n \, g( \mathrm{i} \,\omega_{n}),
\end{equation} 
where $g$ is an arbitrary function.  

The frequency sums encountered below are typically evaluated using the
identity \cite{FW}
\begin{equation} \label{fun:idb}
\frac{1}{\beta} \sum_{n=-\infty}^\infty \frac{1}{-\mathrm{i} \hbar
\omega_n + z} =\frac{1}{2} + n(z) ,
\end{equation} 
or because of antisymmetry \cite{DolanJackiw} 
\begin{equation}
\label{eB} 
\frac{1}{\beta} \sum_{n=-\infty}^\infty \frac{z}{\hbar^2
\omega^2_n + z^2} =\frac{1}{2} + n(z) ,
\end{equation} 
where $n(z)$ is the Bose-Einstein distribution function 
\begin{equation} 
\label{fun:be} 
n(z) = \frac{1}{\mathrm{e}^{\beta z} -1} = \frac{1}{\beta z} - \frac{1}{2} +
\mathcal{O}(\beta z).
\end{equation} 
The first term at the right hand of the identity (\ref{fun:idb}) gives
the result for zero temperature, while the second gives the thermal
contribution.  Such a splitting is convenient because the pure quantum
$T=0$ part can often be combined with the bare (zero-loop) theory.  In a
renormalizable theory, for example, the zero-temperature part leads to a
renormalization of the parameters of the bare theory.  These
contributions are then combined with the bare theory to dress its
parameters.

In this paper, the thermal contributions to the grand canonical thermodynamic
potential $\Omega$ of various systems are expanded in a high-temperature
series.  According to standard statistical mechanics, $\Omega$ is related to
the partition function $Z$ via \cite{Huang}
\begin{equation} 
\Omega = - \beta^{-1} \ln(Z).
\end{equation} 
Physically, this potential determines the thermodynamic pressure $p$
through $\Omega = -p V$, with $V$ the volume of the system.
High-temperature expansions are an important computational tool for
studying interacting theories where usually no exact results can be
derived as for free theories.  The standard approach is to use the
identity (\ref{fun:idb}) backwards and to replace the Bose-Einstein
distribution function again with an expression involving a frequency
sum:
\begin{equation} 
\label{Laurent}
n(z) = \frac{1}{\beta z} -\frac{1}{2} + \frac{2}{\beta} \sum_{n=1}^\infty
\frac{z}{\hbar^2 \omega_n^2 + z^2},
\end{equation} 
where in the first term at the right hand, the $n=0$ contribution is
isolated.  The second term cancels the $T=0$ contribution contained in the
sum.  These two terms correspond to the first terms in a Laurent expansion
of the Bose-Einstein distribution function.

The $n=0$ contribution, stemming from the $\phi_0(\mathbf{k})$ mode in
the Fourier series (\ref{fs}), is of special interest to us.  This
zero-frequency mode, specific to bosonic systems, is a thermal mode for
which the temperature-induced energy gap $2\pi \beta^{-1} n$ in the
denominator at the left hand of Eq.~(\ref{fun:idb}) vanishes.  It is
related to the static field $\Phi$ alluded to above as follows:
\begin{equation}
\label{op} 
\Phi(\mathbf{x}) = \int_0^{\hbar \beta} \mathrm{d} \tau \,
\phi(\tau,\mathbf{x}) = \int \frac{\mathrm{d}^d k}{(2\pi)^d} \,
\mathrm{e}^{\mathrm{i} \mathbf{k} \cdot \mathbf{x}} \, \phi_0 (\mathbf{k}),
\end{equation} 
where use is made of the Fourier series (\ref{fs}).  This classical,
zero-frequency mode is the order parameter of BEC in a weakly
interacting Bose gas.  It develops a nonzero value in the ordered state
where the U(1) symmetry $\phi \to \exp (\mathrm{i} \theta) \phi$
($\theta$ being a constant transformation parameter) is spontaneously
broken, with $|\Phi|^2$ determining the average number density of
condensed particles.  That is, a nonzero value of the order parameter (a
classical object) implies the presence of a condensate (a quantum
phenomenon).  Being static, this mode is unaware of the time dimension
and only notes the space dimensions.  Compared to the dynamic modes
$\phi_n(\mathbf{k})$, with $n \neq 0$, it lives in one dimension less
and its contributions are accordingly different in form from those of
the gapped thermal modes, being typical for a $d$-dimensional rather
than a ($d+1$)-dimensional theory.

The following hierarchy can be identified, with the pure quantum
zero-temperature modes living in $d+1$ noncompact spacetime dimensions, the
gapped thermal modes also living in $d+1$ spacetime dimensions, but with a
compactified time dimension, and finally the zero-frequency mode living in
$d$ space dimensions, which can be understood as ($d+1$)-dimensional
spacetime with the compact time dimension shrank to a point by letting
$\beta \to 0$.  Below, the zero-frequency mode contributions are labeled by
a superscript $0$ to indicate that they correspond to $\beta=0$, while the
zero-temperature ($\beta = \infty$) contributions are labeled by a
superscript $\infty$.  The contributions described by the remaining terms in
the Laurent expansion (\ref{Laurent}), finally, are labeled by a superscript
$\beta$.  We will, somewhat loosely, refer to these last contributions as
due to the gapped thermal modes, although they also encode the $T=0$
contributions, that are explicitly subtracted by the second term in the
Laurent expansion.

Zero-frequency modes can produce nonanalytic behavior, depending on the
number of space dimensions.  Dolan and Jackiw \cite{DolanJackiw} in their
study of thermal phase transitions in relativistic quantum systems took a
pragmatic approach toward these nonanalytic terms induced by such modes and
simply discarded them when estimating critical temperatures.  In light of
the discussion above, this approach is justified as the critical temperature
is to first order determined by the vanishing of the quadratic term in the
effective theory \cite{RR}.  Featuring the zero-frequency mode $\Phi$, that
theory is obtained by integrating out the gapped thermal modes, which give
only analytic contributions.  It is only in the next stage, where
fluctuations in $\Phi$ are studied, that nonanalytic contributions can
appear.

\subsection{Outline}
In this paper, regularization techniques routinely applied to handle
ultraviolet divergent integrals are used to stem \textit{infrared}
divergences instead.  Usually, such divergences are regularized by
introducing an infrared cutoff.  This type of regularization forms the
infrared counterpart of the momentum cutoff regularization of
ultraviolet divergences, where integrals are made finite by integrating
only up to a finite momentum.  Besides using momentum cutoff
regularization, ultraviolet divergences can also be rendered finite by
using dimensional regularization, where the integrals are generalized to
arbitrary dimension, or analytic regularization, where powers appearing
in the integrands are generalized to take arbitrary values.  In the
following, these techniques rather than introducing an infrared cutoff
are used to regularize infrared divergences.

The infrared divergences in this paper arise when expanding
thermodynamic potentials in a high-temperature series, using the
algorithm of Ref.~\cite{effbos}.  The algorithm differs from the
standard high-temperature expansion and focuses on the analytic terms
needed for the study of the critical properties of an interacting
system.  A series of examples is given to show in detail that this
high-temperature expansion in conjunction with the regularization
techniques used is reliable, by demonstrating that known results are
recovered.  The examples include a nonrelativistic (Secs.~\ref{sec:BEC}
and \ref{sec:highT}) and a relativistic (Sec.~\ref{sec:RBG}) free Bose
gas, as well as their fermionic counterparts
(Secs.~\ref{sec:NFG}--\ref{sec:RFG}), and a BCS superconductor
(Sec.~\ref{sec:BCS}).  In Sec.~\ref{sec:Tc}, fluctuations in the nonzero
frequency modes of a weakly interacting Bose gas are considered by
integrating out these gapped thermal modes in a loop expansion.  In
particular their effect on the condensation temperature is studied using
the algorithm advocated here.  As mentioned at the beginning of the
Introduction, although irrelevant for the critical behavior itself,
these high-energy modes can in principle affect the value of the
condensation temperature, which--unlike the critical exponents--is
nonuniversal and depends on the details of the underlying 
quantum theory.  Owing to the temperature-induced energy gaps, the
contributions due to the nonzero frequency modes can be calculated in
perturbation theory.  It is shown that the shift in the condensation
temperature found in Ref.~\cite{effbos} at one loop is canceled by
two-loop contributions \cite{Arnold}, not only in $d=3$ but in arbitrary
$2 < d <4$ (Sec.~\ref{sec:generald}).  The advantage of considering
arbitrary dimensions is that BEC in two dimensions can be studied by
taking the limit $d \to 2$.  A discussion of some recent studies on the
subject is given in Sec.~\ref{sec:comp} followed by conclusions in the
last section (Sec.~\ref{sec:concl}).

\section{Bosonic Integrals}

\subsection{Bose-Einstein Condensation}
\label{sec:BEC}
As first bosonic system, a free Bose gas \textit{above} the condensation
temperature is considered.  Being noninteracting, the field $\phi_n
(\mathbf{k})$ describing the bosons can be integrated out exactly, leading
to the thermodynamic potential $\Omega$ \cite{Huang} 
\begin{equation} 
\label{explicit}
\frac{\Omega}{V} = \beta^{-1} \sum_{n=-\infty}^\infty \int
\frac{\mathrm{d}^d k}{(2\pi)^d} \ln \left(- \mathrm{i} \hbar\omega_n +
\frac{\hbar^2 k^2}{2m} - \mu\right) ,
\end{equation} 
apart from irrelevant constant terms independent of $\beta$ and $\mu$.
Here, $m$ and $\lambda$ are the mass and the de Broglie thermal
wavelength $\lambda = \sqrt{2 \pi \hbar^2 \beta/m}$ of the particles,
and $\mu<0$ is the chemical potential.  The argument of the logarithm is
recognized as the kernel of the time-dependent Schr\"odinger equation in
the imaginary-time formalism, describing a particle of mass $m$ at
finite temperature in a constant background potential $-\mu$.  As
indicated by the single set of frequency sum and momentum integrals,
Eq.~(\ref{explicit}) is a one-loop result that, because the system is
noninteracting, is exact.  The momentum integral is best carried out by
using the Schwinger propertime representation of the logarithm and
Poisson's summation formula
\begin{equation} 
\label{Poisson}
\sum_{n=-\infty}^\infty {\rm e}^{2\pi \mathrm{i} n a} =
\sum_{w=-\infty}^\infty \delta(a-w)
\end{equation}
to replace the summation over $n$ by one over $w$.  The integrals over the
Schwinger propertime parameter and the momentum variable are then easily
carried out to yield the familiar fugacity series \cite{Huang}
\begin{equation} 
\label{loops}
\frac{\Omega}{V} = - \frac{1}{\beta \lambda^d} \sum_{w=1}^\infty w^{- \tau}
\mathrm{e}^{-\alpha w },
\end{equation} 
where $\tau = d/2 +1$ and $\alpha = -\beta \mu>0$.  In Eq.~(\ref{loops}),
the singular zero-temperature contribution corresponding to $w=0$ is
omitted.  A more careful derivation of the thermodynamic potential
(\ref{explicit}) reveals the presence of an extra convergence factor
$\exp(\mathrm{i} \omega_n \eta)$ typical for nonrelativistic theories, where
$\eta$ is a small positive parameter that is taken to zero only after the
frequency sum is evaluated \cite{FW}.  This convergence factor suppresses
the $T=0$ contribution (see below).  Physically, the series (\ref{loops})
denotes the sum over closed worldlines that wind $w$ times around the
imaginary time axis \cite{loops}.  The first factor in the summand measures
the configurational entropy of the loops, while the second, with $\alpha$
being proportional to the worldline tension, is a Boltzmann factor, weighing
loops according to their length measured by $w$.  Loops with $w>1$
correspond to exchange rings of $w$ particles which are cyclically permuted
after an imaginary time $\hbar \beta$, as they appear in Feynman's theory of
the lambda transition in superfluid $^4$He \cite{lambda}.

The expression (\ref{explicit}) can alternatively be evaluated by
keeping the momentum integral to the end and instead of the Poisson
formula using the identity (\ref{fun:idb}).  For nonrelativistic free
theories, the summand there includes the convergence factor
$\exp(\mathrm{i} \omega_n \eta)$ that suppresses the $T=0$ part
\cite{FW}.  The thermodynamic potential then takes the well-known form
\cite{Huang}
\begin{equation} 
\label{Fint}
\frac{\Omega}{V}  = - \frac{1}{\beta \lambda^d} F_{d/2+1} (\alpha),
\end{equation} 
where
\begin{equation}
\label{F} 
F_\tau(\alpha) = \frac{2}{\Gamma(\tau)} \int_0^\infty \mathrm{d} q
\frac{q^{2 \tau-1}}{\mathrm{e}^{q^2+\alpha}-1}
\end{equation} 
with $q$ the dimensionless loop variable defined by $q^2 = \beta \hbar^2
k^2/2m$.

On comparing the two alternative evaluations of the thermodynamic potential
it follows that
\begin{equation} 
\label{BE}
\sum_{w=1}^\infty w^{- \tau} \mathrm{e}^{-\alpha w } =
\frac{2}{\Gamma(\tau)} \int_0^\infty \mathrm{d} q
\frac{q^{2 \tau-1}}{\mathrm{e}^{q^2+\alpha}-1},
\end{equation} 
where in the first representation the momentum integral has been carried
out, while in the second it still has to be performed.  The equivalence of
both results is easily established by expanding the integrand in a geometric
series and using the integral representation of the Gamma function.

\subsection{High-Temperature Expansion}
\label{sec:highT}
Next, the function $F_\tau(\alpha)$ is expanded in a high-temperature
series.  As mentioned in the Introduction, such an expansion is an
important computational tool for interacting theories, where exact
results are absent.  The standard approach is to use the identity
(\ref{fun:idb}) backwards, i.e., the Laurent expansion (\ref{Laurent})
of the Bose-Einstein distribution function.  The zero-frequency mode,
corresponding to the first term in the Laurent expansion, gives the
contribution
\begin{equation} 
\label{FI}
F_\tau^0(\alpha) = \frac{2}{\Gamma(\tau)} \int_0^\infty
\mathrm{d} q \frac{q^{2 \tau-1}}{q^2 + \alpha} = \alpha^{\tau -1}
\Gamma(1 - \tau),
\end{equation} 
which for $d=3$ leads to a nonanalytic contribution of the form $\sim
\alpha^{3/2}$.  

As an aside, this contribution due to the zero-frequency mode can also
be obtained from the worldline loop representation by replacing the sum
over $w$ in $F_\tau(\alpha)$ by an integral:
\begin{equation} 
\sum_{w=1}^\infty w^{- \tau} \mathrm{e}^{-\alpha w } \to \int_0^\infty
\mathrm{d} w \, w^{- \tau} \mathrm{e}^{-\alpha w } = \alpha^{\tau -1}
\Gamma(1 - \tau) = F_\tau^0(\alpha). 
\end{equation} 

The zero-frequency mode $\phi_0(\mathbf{k})$ is important for our purposes
as it determines the critical properties of the free Bose gas.  To
illustrate this, we differentiate the zero-frequency mode contribution
twice with respect to the chemical potential $\mu$ to obtain the known
result that the compressibility $\kappa$ of an ideal Bose gas diverges as
$\kappa \sim (-\mu)^{\tau-3}$ when the condensation temperature $T_0$ is
approached from above, where $\mu$ vanishes as $\mu \sim - (T -
T_0)^{2/(d-2)}$.  For $d=3$, this implies that
\begin{equation} 
\label{compr}
\kappa \propto (-\mu)^{-1/2} \sim (T -T_0)^{-1},
\end{equation} 
and gives as value for the specific-heat critical exponent
$\alpha_\mathrm{ce} = -1$ \cite{GuntonBuckingham}.  Describing a free
theory, this one-loop result is exact.

Despite being noninteracting, a free Bose gas is not in the universality
class of the Gaussian model, but in that of the spherical model
\cite{GuntonBuckingham}.  The nontrival exponents derive from the constraint
that the total number of particles be fixed.  Without this constraint, the
chemical potential would vanish instead as $\mu \sim T_0 -T$ irrespective of
the dimension.  This would result in Gaussian exponents describing BEC in a
free Bose gas at constant pressure \cite{GuntonBuckingham}.  In four space
dimensions, corresponding to the upper critical dimension, and above (i.e.,
$d \geq 4$), the exponents of the spherical and Gaussian model coincide.

As mentioned before, the zero-temperature contribution vanishes because
of the presence of a convergence factor in nonrelativistic free
theories.  More in line with the present approach, it vanishes even
without including this factor,
\begin{equation} 
\label{v}
-F_\tau^\infty(\alpha) = \frac{1}{\Gamma(\tau)} \int_0^\infty \mathrm{d} q
q^{2 \tau - 1} = 0,
\end{equation} 
because integrals over polynomials give zero within the dimensional
regularization scheme \cite{diagrammar}.  

The rest of the terms in the high-temperature expansion, corresponding to
the last terms in the Laurent expansion (\ref{Laurent}), are analytic.
These contributions are obtained in the standard approach by expanding the
summand in a high-temperature Taylor series before carrying out the momentum
integral and the frequency sum, with each term giving a Riemann zeta
function
\begin{equation} 
\zeta(s) = \sum_{m=1}^\infty \frac{1}{m^s} .
\end{equation} 
Specifically, one finds \cite{London}
\begin{equation} 
\label{highT}
F_\tau^\beta(\alpha) = \sum_{l=0}^\infty \frac{\zeta(\tau
-l)}{\Gamma(l+1)} (-\alpha)^l,
\end{equation} 
with $\alpha$ the dimensionless expansion parameter.  The condensation
temperature is determined by these contributions due to the gapped thermal
modes.  At that temperature, the chemical potential vanishes, so that the
equation for the particle number density $n$,
\begin{equation} 
n= - \frac{1}{V} \frac{\partial \Omega}{\partial \mu} ,
\end{equation} 
with $F_\tau(\alpha)$ in Eq.~(\ref{Fint}) replaced with $F_\tau^\beta(\alpha)$,
gives the condensation temperature
\begin{equation}
\label{T0d}
k_\mathrm{B} T_0 = \frac{2 \pi \hbar^2}{m} \left(
\frac{n}{\zeta(d/2)} \right)^{2/d}.
\end{equation}
In other words, although the critical exponents of BEC in a free Bose
gas are determined entirely by the static, zero-frequency mode, the
condensation temperature, being nonuniversal, is determined entirely by
the dynamic, gapped thermal modes.

It is natural to ask whether the series (\ref{highT}) can also be extracted
directly from the integral representation given in Eq.~(\ref{F}), where the
frequency sum has already been carried out, without returning to a
representation involving such a sum.  To show that the answer is
affirmative, we expand the integrand there in a high-temperature series.
Each term thus generated contains an integral of the form
\begin{equation} 
\int_0^{\infty} \frac{\mathrm{d} q}{q} \frac{q^{2 t} }{\mathrm{e}^{p q^2} -
1}
\end{equation} 
and derivatives thereof with respect to the parameter $p$ (which is set to
unity at the end).  For $t\leq1$, these loop integrals diverge in the
infrared.  We handle the divergences using zeta function regularization, by
analytically continuing the following equation
\begin{equation} 
\label{ac}
2 \int_0^{\infty} \frac{\mathrm{d} q}{q} \frac{q^{2 t} }{\mathrm{e}^{p
q^2} - 1} = \Gamma(t ) \zeta(t ) p^{-t},
\end{equation} 
with $t$ initially chosen large enough so that the integral converges, to
arbitrary values of $t$.  With this regularization scheme, the
high-temperature expansion (\ref{highT}) easily follows,
\begin{equation} 
\frac{2}{\Gamma(\tau)} \int_0^\infty \mathrm{d} q
\frac{q^{2 \tau-1}}{\mathrm{e}^{q^2+\alpha}-1} \;
\stackrel{\mathrm{Taylor}}{\longrightarrow} \; \sum_{l=0}^\infty
\frac{\zeta(\tau -l)}{\Gamma(l+1)} (-\alpha)^l = F_\tau^\beta(\alpha),
\end{equation} 
showing that the contributions due to the gapped thermal modes can
indeed (when properly regularized) be extracted directly from the
function $F_\tau(\alpha)$ in Eq.~(\ref{F}) without the need to return to
the representation involving a frequency sum.  This result is quite
general and will in the next subsection be illustrated for a
relativistic system.

\subsection{Relativistic Free Bose Gas}
\label{sec:RBG}
To show that the conclusions of the previous section by no means hinge on
the form of the nonrelativistic spectrum, leading to the specific structure
of the integrand in Eq.~(\ref{F}), the same analysis for a relativistic free
Bose gas is given in this section.  The system is considered at zero
chemical potential $\mu=0$.  It has no phase transition then, but the
algebra is simpler than with a chemical potential included
\cite{HaberWeldon}, without losing any of the generalities we wish to point
out.

The analog of expression (\ref{explicit}) for the nonrelativistic Bose
gas reads \cite{Kapusta}
\begin{equation} 
\label{explicitrel}
\frac{\Omega}{V} = \frac{1}{2 \beta} \sum_{n=-\infty}^\infty \int
\frac{\mathrm{d}^d k}{(2\pi)^d} \ln \left(\hbar^2 \omega^2_n + \hbar^2
c^2 k^2 + m^2 c^4 \right),
\end{equation} 
with $c$ the speed of light.  The argument of the logarithm is now
recognized as the kernel of the Klein-Gordon equation in the
imaginary-time formalism, describing a relativistic free scalar particle
of mass $m$ at finite temperature.  Being a noninteracting system, the
one-loop result (\ref{explicitrel}) is exact.  With the help of the
identity (\ref{fun:idb}), the thermodynamic potential becomes
\begin{eqnarray} 
\frac{\Omega}{V} &=& \int \frac{\mathrm{d}^d k}{(2\pi)^d}
\left[\frac{1}{2} \sqrt{\hbar^2 c^2 k^2 + m^2 c^4} + \beta^{-1}
\ln\left( 1 - \mathrm{e}^{-\beta \sqrt{\hbar^2 c^2 k^2 + m^2 c^4 }}
\right) \right] \nonumber \\ &=& \frac{(\hbar c)^{-d} \beta^{-(d+1)}}{(4
\pi)^{d/2}} \left[ \frac{1}{\Gamma(d/2)} \int_0^\infty \mathrm{d} q
q^{d-1} \sqrt{ q^2 + (\beta m c^2)^2} - \frac{\Gamma(2
\tau)}{\Gamma(\tau)} H_{\tau}(\beta m c^2) \right],
\end{eqnarray} 
where the new integration variable is $q = \beta \hbar c k$ and the
thermal part of the thermodynamic potential is determined by the
function
\begin{equation}  
\label{H}
H_{\tau} (y) = \frac{1}{\Gamma(2\tau)} \int_0^\infty \frac{\mathrm{d}
q}{\sqrt{q^2 + y^2}} \frac{q^{2\tau-1}}{\mathrm{e}^{\sqrt{q^2 +
y^2}}-1},
\end{equation} 
with $\tau = d/2 + 1$.  This function \cite{Kapusta} is the relativistic
analog of the function $F_{\tau} (\alpha)$ in Eq.~(\ref{F}).

The high-temperature expansion of the function $H_{\tau} (y)$ in $d=3$
was computed by Dolan and Jackiw \cite{DolanJackiw}.  To deal with
logarithmic divergences that arise because four spacetime dimensions
corresponds to the upper critical dimension, dimensional regularization
was used by considering $d=3-\epsilon$, with $\epsilon$ taken to zero at
the end.  For convenience, $\tau=5/2-\epsilon/2$ was only used in the
integrand, whereas $\epsilon$ was set to zero from the start in the
prefactor in Eq.~(\ref{H}). These authors used Eq.~(\ref{eB}) backwards,
i.e., Eq.~(\ref{Laurent}), expanded the summand in a Taylor series,
integrated term by term, and finally carried out the remaining frequency
sum in each term, with the result \cite{DolanJackiw}:
\begin{equation}
H_{5/2} (y) = \frac{\pi^4}{360} - \frac{\pi^2}{96} y^2 + \frac{\pi}{48} y^3
+ \frac{1}{128} \left[\ln\left(\frac{y}{4 \pi}\right) + \gamma -
\frac{3}{4} \right] y^4 + \mathcal{O}(y^6) ,
\end{equation} 
where $\gamma$ is Euler's constant and $y=\beta m c^2$ the dimensionless
expansion parameter.  The nonanalytic cubic contribution is due to the
zero-frequency mode represented by the first term in the Laurent expansion
of the Bose-Einstein distribution function in Eq.~(\ref{H}):
\begin{equation} 
H_{5/2}^0(y) = \frac{1}{\Gamma(5)} \int_0^\infty \mathrm{d} q
\frac{q^{4}}{q^2 + y^2} = \frac{\pi}{48} y^3 ,
\end{equation} 
where analytic regularization is used to handle the ultraviolet divergence.
The second term in the Laurent expansion of the Bose-Einstein distribution
function gives
\begin{equation}
\label{Hinfty} 
-H_{5/2}^\infty(y) = - \frac{1}{2} \frac{1}{\Gamma(5)} \int_0^\infty
\mathrm{d} q \frac{q^{4-\epsilon}}{\sqrt{q^2 + y^2}} = \frac{1}{128}
\left[ - \frac{1}{\epsilon} + \ln \left(\frac{y}{2} \right) +
\frac{7}{12} \right] y^4,
\end{equation} 
while the gapped thermal modes yield
\begin{equation} 
H_{5/2}^\beta(y) = \frac{\pi^4}{360} - \frac{\pi^2}{96} y^2 +
\frac{1}{128} \left[\frac{1}{\epsilon} - \ln (2 \pi) + \gamma -
\frac{4}{3} \right] y^4 + \mathcal{O}(y^6) ,
\end{equation} 
so that $H_{5/2}(y) = H_{5/2}^0(y) - H_{5/2}^\infty(y) + H_{5/2}^\beta(y)$.
An extra minus sign is included in the left hand of Eq.~(\ref{Hinfty}), so that
$H_{5/2}^0$ is the $T=0$ contribution that has to be subtracted from
$H_\tau^\beta(\alpha)$ to arrive at the purely thermal contributions
$H_\tau(\alpha)$.

We next rederive this result using the algorithm given above and extract it
directly from the integral representation (\ref{H}), by expanding the
integrand in a Taylor series.  Applying zeta function regularization to
handle the infrared divergences as before [see Eq.~(\ref{ac})], we obtain
\begin{eqnarray} 
\label{Hal}
\frac{1}{\Gamma(5)} \int_0^\infty \frac{\mathrm{d} q}{\sqrt{q^2 + y^2}}
\frac{q^{4 - \epsilon}}{\mathrm{e}^{\sqrt{q^2 + y^2}}-1}
\stackrel{\mathrm{Taylor}}{\longrightarrow} && \frac{\pi^4}{360} -
\frac{\pi^2}{96} y^2 + \frac{1}{128} \left[ \frac{1}{\epsilon} + 2
\zeta'(0) + \gamma - \frac{4}{3} \right] y^4 \nonumber \\ && +
\frac{\zeta'(-2)}{768} y^6 + \frac{\zeta'(-4)}{24576} y^8 +
\frac{\zeta'(-6)}{1474560} y^{10} + \mathcal{O}(y^{12}) ,
\end{eqnarray} 
where this series is carried out to higher order.  To show that this
expression coincides with $H_{5/2}^\beta(y)$, we differentiate the
identity \cite{AS}
\begin{equation} 
\zeta(s) = 2^s \pi^{s-1} \sin \left(\tfrac{1}{2} \pi s \right)
\Gamma(1-s) \zeta(1-s),
\end{equation} 
to obtain for $s$ a (positive or negative) even integer:
\begin{equation} 
\zeta'(s) = 2^{s-1} \pi^{s} \cos \left(\tfrac{1}{2} \pi s \right)
\Gamma(1-s) \zeta(1-s),
\end{equation} 
or specifically
\begin{equation} 
\label{zetad}
\zeta'(0) = - \tfrac{1}{2} \ln(2 \pi),  \quad \zeta'(-2) = - \frac{1}{4
\pi^2} \zeta(3), \quad \zeta'(-4) = \frac{3}{4 \pi^4} \zeta(5), \quad
\zeta'(-6) = - \frac{45}{8 \pi^6} \zeta(7).
\end{equation} 
It now immediately follows that Eq.~(\ref{Hal}) gives the contributions
due to the gapped thermal modes, so that the results obtained in the
standard manner and the one used here indeed coincide.

Given that also the zero-frequency mode contribution as well as the $T=0$
part can be extracted from the integral representation (\ref{H}), by
identifying them with the first two terms in the Laurent expansion of the
Bose-Einstein distribution function, it follows that no backtracking to a
representation involving a frequency sum is required, provided zeta function
regularization is applied.  Moreover, the simple, yet powerful algorithm
discussed here is easily implemented using an algebraic manipulation program
to generate large orders in a high-temperature expansion for a wide class of
problems, including fermionic systems (with the obvious changes), as will
now be demonstrated.

\section{Fermionic Integrals}

\subsection{Free Fermi Gas}
\label{sec:NFG}
In this subsection, a nonrelativistic free, spinless Fermi gas is
considered.  The thermodynamic potential is given by the one-loop expression
(\ref{explicit}) for a Bose gas with an extra minus sign included to account
for the antisymmetry of the fermionic field.  Using Poisson's summation
formula (\ref{Poisson}), the thermodynamic potential is readily written as a
fugacity series \cite{Huang}
\begin{equation} 
\label{loopsf}
\frac{\Omega}{V} = - \frac{1}{\beta \lambda^d} \sum_{w=1}^\infty
(-1)^{w+1}w^{- \tau} \mathrm{e}^{-\alpha w },
\end{equation} 
where $\tau = d/2 +1$ and $\alpha = -\beta \mu>0$ again.  As for a Bose
gas, this series can be understood as representing a sum over closed
worldlines winding around the imaginary time axis $w$ times, with each
winding accompanied by a factor $-1$ typical for fermions.

Alternatively, the thermodynamic potential can be evaluated using the identity
\cite{FW}
\begin{equation} 
\label{fun:idf}
\frac{1}{\beta} \sum_{n=-\infty}^\infty \frac{1}{-\mathrm{i} \hbar
\omega_n + z} = \frac{1}{2} - f(z) ,
\end{equation} 
where $\omega_n = \pi \beta^{-1} (2n + 1) $ are the fermionic Matsubara
frequencies and $f(z)$ is the Fermi-Dirac distribution function
\begin{equation} 
\label{fun:fd} 
f(z) = \frac{1}{\mathrm{e}^{\beta z} +1} = \frac{1}{2} + \mathcal{O}(\beta z).
\end{equation} 
The first term at the right hand of the identity (\ref{fun:idf}) gives
the result for zero temperature, while the second gives the thermal
contribution.  Nonrelativistic free theories include again a convergence
factor which suppresses the $T=0$ part.  The thermodynamic potential then
takes the same form (\ref{Fint}) as for a Bose gas with the function
$F_\tau(\alpha)$ replaced with the function
\begin{equation}
\label{G} 
G_\tau(\alpha) = \frac{2}{\Gamma(\tau)} \int_0^\infty \mathrm{d} q
\frac{q^{2 \tau-1}}{\mathrm{e}^{q^2+\alpha}+1}.
\end{equation} 
The equivalence of the two representations, implying 
\begin{equation}
\label{equi} 
\sum_{w=1}^\infty (-1)^{w+1}w^{- \tau} \mathrm{e}^{-\alpha w } =
\frac{2}{\Gamma(\tau)} \int_0^\infty \mathrm{d} q \frac{q^{2
\tau-1}}{\mathrm{e}^{q^2+\alpha}+1},
\end{equation} 
is easily established.

\subsection{High-Temperature Expansion}
As for a bosonic system, the standard high-temperature series expansion of
the function $G_\tau(\alpha)$ is to use the identity (\ref{fun:idf})
backwards and to replace the Fermi-Dirac distribution function again with an
expression involving a frequency sum:
\begin{equation} 
\label{Laurentf}
f(z) = \frac{1}{2} - \frac{2}{\beta} \sum_{n=0}^\infty
\frac{z}{\hbar^2 \omega_n^2 + z^2},
\end{equation} 
where the first term at the right hand subtracts the $T=0$ part encoded in
the sum.  Unlike bosonic systems, fermionic systems do not have a
zero-frequency mode as none of the Matsubara frequencies (\ref{bcs:Mats})
vanishes.

Following the algorithm advocated here, we instead derive this expansion
directly from the integral representation of the function $G_\tau(\alpha)$
given in Eq.~(\ref{G}) without first returning to an expression involving a
frequency sum.  Expanding the integrand in Eq.~(\ref{G}) in a Taylor series,
we encounter loop integrals of the form
\begin{equation} 
\int_0^{\infty} \frac{\mathrm{d} q}{q} \frac{q^{2t} }{\mathrm{e}^{p
q^2} + 1},
\end{equation}  
which for $t \leq 0$ are infrared divergent.  These divergences are handled
again using zeta function regularization, where the following equation
\begin{equation} 
2 \int_0^{\infty} \frac{\mathrm{d} q}{q} \frac{q^{2t} }{\mathrm{e}^{p
q^2} + 1} = \Gamma(t) \left(1 - 2^{1 - t }\right) \zeta(t ) p^{-t},
\end{equation}  
with $t$ initially chosen large enough so that the integral converges, is
analytically continued to arbitrary values of $t$.  The high-temperature
expansion now follows as: 
\begin{equation} 
\frac{2}{\Gamma(\tau)} \int_0^\infty \mathrm{d} q \frac{q^{2
\tau-1}}{\mathrm{e}^{q^2+\alpha}+1}
\stackrel{\mathrm{Taylor}}{\longrightarrow} \; \sum_{l=0}^\infty \left(1
- 2^{1-\tau + l} \right) \frac{\zeta(\tau -l)}{\Gamma(l+1)} (-\alpha)^l =
G_\tau^\beta(\alpha) .
\end{equation} 
This result can be checked by substituting Eq.~(\ref{equi}) and
differentiating the resulting equation arbitrary many times with respect
to the dimensionless expansion parameter $\alpha$ and setting that
parameter to zero at the end, leading to
\begin{equation} 
\sum_{w=1}^\infty (-1)^{w+1}w^{-t} = \left(1 - 2^{1 - t}\right)
\zeta(t),
\end{equation} 
which is an identity.

\subsection{Relativistic Free Fermi Gas}
\label{sec:RFG}
As last noninteracting example, a relativistic free Fermi gas at zero
chemical potential is considered, consisting of particles (but no
antiparticles, which would lead to an extra degeneracy factor) with
half-integer spin $\sigma$.  The thermodynamic potential is given by the
expression (\ref{explicitrel}) for the Bose gas with an extra spin
multiplication factor and a minus sign, i.e., $-(2 \sigma +1)$,
included.  With $\omega_n$ now denoting fermionic Matsubara frequencies,
the one-loop expression is readily cast in the form
\begin{equation} 
\frac{\Omega}{V} = - (2\sigma+1) \frac{(\hbar c)^{-d} \beta^{-(d+1)}}{(4
\pi)^{d/2}} \left[ \frac{1}{\Gamma(d/2)} \int_0^\infty \mathrm{d} q q^{d-1}
\sqrt{ q^2 + (\beta m c^2)^2} - \frac{\Gamma(2 \tau)}{\Gamma(\tau)}
I_{\tau}(\beta m c^2) \right],
\end{equation} 
where the thermal part is determined by the function
\begin{equation}  
\label{I}
I_{\tau} (y) = \frac{1}{\Gamma(2\tau)} \int_0^\infty \frac{\mathrm{d}
q}{\sqrt{q^2 + y^2}} \frac{q^{2\tau-1}}{\mathrm{e}^{\sqrt{q^2 +
y^2}}+1}.
\end{equation} 
For $\tau=5/2$, it has the high-temperature expansion \cite{DolanJackiw}
\begin{equation} 
\label{I52}
I_{5/2} (y) = \frac{7  \pi^4}{2880} - \frac{\pi^2}{192} y^2 -
\frac{1}{128} \left[\ln\left(\frac{y}{\pi} \right) + \gamma - \frac{3}{4}
\right] y^4 + \mathcal{O}(y^6) ,
\end{equation} 
where because of the absence of a zero-frequency mode, all terms
are analytic in the dimensionless expansion parameter $y=\beta m c^2$.

Apart from a minus sign, the $T=0$ contribution is the same as that for a
Bose gas given in Eq.~(\ref{Hinfty}):
\begin{equation} 
-I_{5/2}^\infty(y) =  \frac{1}{2} \frac{1}{\Gamma(5)} \int_0^\infty
\mathrm{d} q \frac{q^{4-\epsilon}}{\sqrt{q^2 + y^2}} = \frac{1}{128}
\left[\frac{1}{\epsilon} - \ln \left(\frac{y}{2} \right) -
\frac{7}{12} \right] y^4,
\end{equation} 
where again an extra minus sign is included in the left hand of this
definition, so that $I_{5/2}^0$ gives the $T=0$ contribution (and not its
negative).

Using the algorithm discussed here, we expand the integrand of
Eq.~(\ref{I}) in a Taylor series to obtain as contribution due to the
gapped thermal modes: 
\begin{eqnarray} 
\frac{1}{\Gamma(2\tau)} \int_0^\infty \frac{\mathrm{d} q}{\sqrt{q^2 +
y^2}} \frac{q^{2\tau-1}}{\mathrm{e}^{\sqrt{q^2 + y^2}}+1}
\stackrel{\mathrm{Taylor}}{\longrightarrow} && \frac{7 \pi^4}{2880} -
\frac{\pi^2}{192} y^2 + \frac{1}{128} \left[ - \frac{1}{\epsilon} -2
\zeta'(0) - \gamma - 2 \ln (2) + \frac{4}{3} \right] y^4 \nonumber \\ &&
- \frac{7}{768} \zeta'(-2) y^6 - \frac{31}{24576} \zeta'(-4) y^8 -
\frac{127}{1474560} \zeta'(-6) y^{10} + \mathcal{O}(y^{12}),
\end{eqnarray} 
which, with the help of the identities (\ref{zetad}) and together with
$I_{5/2}^\infty(y)$, reproduces the result (\ref{I52}) (expanded to
higher order).

These last two examples illustrate that the high-temperature expansion we
use in conjunction with zeta function regularization of infrared divergences
is also valid for fermionic systems.

\subsection{High-Temperature Expansion of BCS Theory}
\label{sec:BCS}
In Ref.~\cite{effbos}, the BCS theory was considered at finite temperature
using the approach advocated here.  This example is included to point out a
shift in approach when going from a noninteracting to an interacting system.
The main difference with a noninteracting system is of course that one-loop
results no longer are exact, meaning that not all the degrees of freedom can
be integrated out in an exact manner to obtain the thermodynamic potential
as a known function of the thermodynamic parameters $\alpha$ and $\beta$.
Instead, irrelevant degrees of freedom are integrated out (usually only
perturbatively) to arrive at an effective theory of the Landau form
expressed in terms of an order parameter describing the phase transition.
The order parameter, whose value is zero in the symmetric, disordered state
and nonzero in the ordered state with spontaneously broken symmetry,
determines the critical properties of the system close to the phase
transition, where it disappears. 

After linearizing the quartic interaction of the BCS theory by a suitable
Hubbard-Stratonovich transformation, the fermionic degrees of freedom can be
integrated out exactly in a one-loop calculation to yield for a uniform BCS
superconductor:
\begin{equation}  
\label{omegabcs} 
\frac{\Omega_1(\Delta)}{V} = - 2 \int \frac{\mathrm{d}^3 k}{(2\pi)^3}
\left[\frac{1}{2} E(k) + \beta^{-1} \ln \left(1 + {\rm e}^{-\beta E(k)}
\right) \right].
\end{equation}
Here, the frequency sum has been carried out and
\begin{equation} 
\label{bcsspec}
E(k) =\sqrt{\xi^2(k) + |\Delta|^2},
\end{equation} 
is the BCS spectrum with the order parameter $\Delta$ providing an energy
gap at the Fermi surface.  This spectrum replaces the excitation spectrum
$\xi(k) = \hbar^2 k^2/2m - \mu$ of the elementary fermionic excitations of
mass $m$ in the normal state, where the energy is measured relative to the
chemical potential $\mu$, which, contrary to a dilute free Fermi gas, is
positive in the weak-coupling BCS limit and given by the Fermi
energy $\mu = \hbar^2 k_\mathrm{F}^2/2m$, with $\hbar k_\mathrm{F}$ the
Fermi momentum.  The prefactor $2$ in Eq.~(\ref{omegabcs}) arises because
the fermions come in two species, with spin up and down.  The true grand
canonical potential $\Omega$, depending only on the thermodynamic
parameters $\alpha$ and $\beta$, is obtained from $\Omega(\Delta)$ by
integrating out $\Delta$. This can only be done perturbatively, with the
simplest approximation corresponding to the saddle point of the integral,
where the order parameter is considered to be a nonfluctuating field.

To the one-loop expression (\ref{omegabcs}), the tree contribution is to be
added
\begin{equation}  
\label{ep0}
\frac{\Omega_0(\Delta)}{V} = - \frac{|\Delta|^2}{g_\mathrm{B} },
\end{equation}
where $g_\mathrm{B} < 0$ is the (bare) coupling constant of the local BCS
interaction term, representing the effective attraction between fermions.
The pure quantum zero-temperature term in Eq.~(\ref{omegabcs}) gives, among
other contributions, a quadratic term which can be combined with the tree
contributions to yield the renormalized coupling constant $g$,
\begin{equation} \label{bcs:reng}
\frac{1}{g} = \frac{1}{g_\mathrm{B}} + \frac{1}{2} \int
\frac{\mathrm{d}^3 k}{(2\pi)^3} \frac{1}{|\xi(k)|}.
\end{equation} 
The integral diverges in the ultraviolet, to regularize it we introduce
a momentum cutoff $\hbar \Lambda$ to obtain
\begin{equation} \label{bcs:ren}
\frac{1}{g} = \frac{1}{g_\mathrm{B}} + \frac{m}{2 \pi^2 \hbar^2} \Lambda +
\mathcal{O}(\Lambda^0),
\end{equation} 
where the irrelevant finite part of the integral is omitted.  The
weak-coupling BCS limit corresponds to taking $g_\mathrm{B} \rightarrow 0$ from
below.

With the loop integration approximated by 
\begin{equation} 
\int \frac{\mathrm{d}^3 k}{(2\pi)^3} \to \nu(0) \int_{-\infty}^\infty
\mathrm{d} \xi,
\end{equation} 
where $\nu(0) = m k_\mathrm{F} / 2 \pi^2 \hbar^2$ is the density of
states per spin degree of freedom at the Fermi level, the thermal part
of the thermodynamic potential (\ref{omegabcs}) becomes of a form
encountered in a relativistic free Fermi gas \cite{DolanJackiw}, with
the dimensionless integration variable $q=\beta \xi$.  The approximation
consists of extending the range of the $\xi$ integration from $-\mu \leq
\xi < \infty$ to $-\infty < \xi < \infty$.  The problem thus reduces to
one in a single space dimension ($d=1$), corresponding to the value
$\tau=3/2$.  To handle logarithmic divergences that arise, we again
dimensional regularize the integral and consider the problem in $1 -
\epsilon$ dimensions instead.  We proceed in the same way as before and
rather than returning to an expression involving a frequency sum by
using Eq.~(\ref{fun:idf}) backwards, we expand the integrand in
Eq.~(\ref{H}) in a high-temperature series and apply zeta function
regularization to handle infrared divergences.  This gives as
contributions due to the gapped thermal modes: \cite{effbos}
\begin{equation}    
\label{hight} 
\int \frac{\mathrm{d}^3 k}{(2\pi)^3} \ln \left(1 + {\rm e}^{-\beta
\sqrt{\xi^2(k) + |\Delta|^2}} \right) = \beta^{-1} \nu(0)
\left\{\frac{\pi^2}{6} + \frac{1}{2} \left[\frac{1}{\epsilon } + 2
\zeta'(0) + \gamma + 2 \ln(2) \right] y^2 + \frac{7}{8} \zeta'(-2) y^4 +
\mathcal{O}(y^6) \right\}.
\end{equation} 
Because the order parameter appears only in the dimensionless expansion
parameter $y=\beta |\Delta|$, the high-temperature expansion is tantamount
to a Landau expansion in $|\Delta|$.  Contact with the standard approach is
established through the identities (\ref{zetad}) and the substitution
\begin{equation}
\label{connect}
\frac{1}{\epsilon } \rightarrow \ln \left(\beta \hbar \omega_\mathrm{D} \right),
\end{equation}
where the Debeye energy $\hbar \omega_\mathrm{D}$, being a measure of
the inverse lattice spacing, is the physical ultraviolet cutoff and the
temperature $\beta^{-1}$ is the relevant infrared scale.  This
correspondence between the pole $1/\epsilon$ of dimensional
regularization and the logarithm $\ln(\Lambda)$ appearing in momentum
cutoff regularization is commonly used in the context of quantum field
theory, and can be verified explicitly here by working with the Debeye
energy as a cutoff from the beginning instead of using dimensional
regularization.  Adding the tree and the one-loop contributions, we then
find
\begin{equation} 
\label{go}
\frac{\Omega_0(\Delta) + \Omega_1(\Delta)}{V} = - \beta^{-2} \nu(0)
\left\{ \frac{\pi^2}{3} + \left[\ln \left(\beta \hbar \omega_\mathrm{D}
\right) + \gamma - \ln (\pi/2 ) + \frac{1}{\nu(0)
g}\right] |\beta \Delta|^2 - \frac{7\pi^2}{16} \zeta(3) |\beta \Delta|^4
\right\} ,
\end{equation} 
featuring the renormalized coupling constant $g$.  Being a static field,
the order parameter is the zero-frequency mode of the problem at hand.
The critical temperature $T_{\rm c}$ is determined by the condition that
the coefficient of the quadratic term changes sign, yielding the
standard result \cite{AGD}
\begin{equation}
\label{TBCS}
k_\mathrm{B} T_{\rm c} = \frac{2}{\pi} {\rm e}^{\gamma } \hbar
\omega_{\rm D} \, {\rm e}^{1/\nu(0) g} ,
\end{equation}
with $g<0$.  Using this expression for the critical temperature,
we can put Eq.~(\ref{go}) in the canonical form \cite{Gorkov}
\begin{equation}
\frac{\Omega_0(\Delta) + \Omega_1(\Delta)}{V} = - \beta^{-2} \nu(0) \left[
\frac{\pi^2}{3} - \ln \left(\frac{\beta_\mathrm{c}}{\beta} \right) |\beta
\Delta|^2 - \frac{7\pi^2}{16} \zeta(3) |\beta \Delta|^4 \right],
\end{equation} 
valid close to the critical temperature.  

This illustrates first of all that our method of computing the contributions
due to the gapped thermal modes in a high-temperature series expansion in
conjunction with zeta function regularization of infrared divergences is
also valid for an interacting system.  In addition, it highlights the
general approach to critical phenomena, which is to integrate out the
irrelevant degrees of freedom (which include the gapped thermal and thus all
fermionic modes) to determine the coefficients of the effective theory
expressed in terms of the order parameter.  Although irrelevant for the
critical behavior itself, the gapped thermal modes can influence the value
of the critical temperature through their contributions to the coefficient
of the quadratic term.  The critical temperature (\ref{TBCS}), which is a
one-loop result, is exclusively determined by the high-energy dynamic modes
as none of the Matsubara frequencies of the fermionic fields that have been
integrated out vanishes.  Unlike critical exponents, the critical
temperature is not universal and can depend on the details of these
high-energy gapped modes.

\section{Condensation Temperature of Weakly Interacting Bose Gas}

\subsection{Effective Theory}
\label{sec:Tc}

We next consider the condensation temperature of a uniform, weakly
interacting Bose gas.  Specifically, we are interested in how
fluctuations in the irrelevant degrees of freedom influence this
temperature.  To determine this, we need, according to the effective
field program (see the discussion in the preceding section), to
calculate the coefficients of the effective theory expressed in terms of
the order parameter by integrating out the irrelevant degrees of
freedom.  For BEC the irrelevant degrees of freedom are the gapped
thermal modes, whose contributions are conveniently computed using the
algorithm advocated here.  The order parameter $\Phi$ of BEC in a weakly
interacting Bose gas, introduced in Eq.~(\ref{op}), denoting the
expectation value of the quantum field $\phi$ describing the bosons, is
the zero-frequency mode.  A nonzero $\Phi$ implies the presence of a
condensate, with $|\Phi|^2$ giving the average number density
$n_\mathrm{vac}$ of condensed particles.  This effective field program
is the \textit{field}-theoretic generalization of the effective
classical potential introduced by Feynman and Kleinert \cite{FK} in
their pathintegral study of thermal fluctuations via a final integral
over the zero-frequency component of the \textit{path} variable.  (See
the textbook \cite{PathI} for extensive applications.)  Here we are left
with a final functional integral over the purely space-dependent field
$\Phi(\mathbf{x})$.

Whereas the condensation temperature of a free Bose gas is determined
from within the normal, i.e., the high-temperature state, we determine
$T_\mathrm{c}$ for the weakly interacting gas instead from within the
low-temperature state with a nonzero condensate $|\Phi|$.

The one-loop contribution to the thermodynamic potential of a weakly
interacting uniform Bose gas is given by \cite{effbos}
\begin{equation} 
\label{bec:Vfinal}
\frac{\Omega_1(\Phi)}{V} = \int \frac{\mathrm{d}^3 k}{(2\pi)^3}
\left[\frac{1}{2} E(k) + \beta^{-1} \ln \left(1 - {\rm e}^{-\beta E(k)}
\right) \right],
\end{equation} 
where $E(k)$ is the spectrum of the elementary excitation in the
superfluid state expressed in terms of the order parameter $\Phi$
\begin{equation} \label{bec:bogog}
E(k) = \sqrt{\left( \hbar^2 k^2/2m  - \mu_\mathrm{B} + 4 g_\mathrm{B}
|\Phi|^2 \right)^2 - 4 g_\mathrm{B}^2 |\Phi|^4 } ,  	
\end{equation}
where $\mu_\mathrm{B}$ and $g_\mathrm{B}$ are the bare parameters of the
theory, with $g_\mathrm{B}> 0$ the coupling constant of the repulsive
two-particle contact interaction.  The true thermodynamic potential
$\Omega(\alpha,\beta)$ is obtained by integrating out the remaining degree
of freedom, which in a first approximation is done by evaluating the
integral in the saddle point, corresponding to extremizing $\Omega(\Phi)$.

Written in the form (\ref{bec:bogog}), the spectrum is the bosonic
analog of the fermionic BCS spectrum (\ref{bcsspec}) given in terms of
the superconducting order parameter $\Delta$.  The spectrum
(\ref{bec:bogog}) reduces to the $T=0$ Bogoliubov spectrum
\cite{Bogoliubov} when the lowest-order value $|\Phi|^2 (=
n_\mathrm{vac}) = \mu_\mathrm{B}/2g_\mathrm{B}$ for $\Phi$ is inserted.
This value is obtained from minimizing the tree contribution
\begin{equation} 
\label{tree}
\frac{\Omega_0(\Phi)}{V} = - \mu_\mathrm{B} |\Phi|^2 + g_\mathrm{B}
|\Phi|^4.
\end{equation} 
It is important to note that the chemical potential $\mu_\mathrm{B} = 2
g_\mathrm{B} n_\mathrm{vac}$ is positive here (so that the
coefficient of the quadratic term is negative, leading to a potential of a
form resembling a Mexican hat), setting a weakly interacting Bose gas apart
from a free Bose gas which has a nonpositive chemical potential.

The pure quantum zero-temperature term in Eq.~(\ref{bec:Vfinal}) gives,
among other contributions, two terms of the form contained in the bare
theory (\ref{tree}).  They can be combined to yield the renormalized
parameters \cite{BBS}:
\begin{eqnarray}
\label{bec:renla}  
\mu &=& \mu_\mathrm{B} - \frac{1}{6\pi^2} g_\mathrm{B} \Lambda^3
\label{bec:renmu} \\ g &=& g_\mathrm{B} - \frac{m}{\pi^2 \hbar^2} g_\mathrm{B}^2
\Lambda,
\end{eqnarray}  
with $\hbar \Lambda$ the momentum cutoff.  In addition, it gives a
nonanalytic contribution $\propto m^{3/2} (g |\Phi|^2)^{5/2}$ induced by the
gapless Goldstone mode.  This one-loop contribution, which is irrelevant for
our present purposes and will subsequently be ignored, is typical for a
gapless contribution in five spacetime dimensions.  The effective number of
spacetime dimensions is five here because in nonrelativistic quantum
theories, where time derivatives appear in combination with two space
derivatives, the time dimension counts double compared to the (three) space
dimensions.

Since Eq.~(\ref{bec:Vfinal}) is already a one-loop result, it is consistent
to this order to replace the bare parameters $\mu_\mathrm{B}, g_\mathrm{B}$
there with the (one-loop) renormalized ones.  The renormalized coupling
constant $g$ is related to the s-wave scattering length $a$ via
\cite{Hugenholtz,AGD}
\begin{equation} \label{bec:a}
g = \frac{2 \pi \hbar^2 a}{m}.
\end{equation} 
The zero-temperature relation between the renormalized, physical chemical
potential and the average number density $n_\mathrm{vac} = |\Phi|^2$ of
condensed particles then reads \cite{FW}
\begin{equation}
\label{mu0} 
\mu(T=0) = 2 g n_\mathrm{vac} = \frac{4 \pi \hbar^2 a}{m} n ,
\end{equation} 
where in the last step we ignored the so-called depletion of the condensate
\cite{Bogoliubov} and replaced $n_\mathrm{vac}$ with the total particle
number density, which is justified because the term is already of order $g$.

Next, the thermal part in Eq.~(\ref{bec:Vfinal}) is expanded in a
high-temperature series using the algorithm discussed above.  Writing
\begin{equation} 
\int \frac{\mathrm{d}^3 k}{(2\pi)^3} \ln \left(1 - {\rm e}^{-\beta E(k)}
\right) = \frac{32 \pi}{(4 \pi)^{3/2}} \frac{1}{\lambda^3} \int_0^\infty
\mathrm{d} q q^2 \ln \left(1 - {\rm e}^{-\beta E(q)} \right),
\end{equation} 
with $q$ the same loop variable as used for a free Bose gas, defined by
$q^2=\beta \hbar^2 k^2/2m$, and
\begin{equation} 
\beta E(q) = \sqrt{\left(q^2 + \alpha + 4 y \right)^2 - 4 y^2},
\end{equation}
where $y= \beta g |\Phi|^2$, we arrive at \cite{effbos}
\begin{eqnarray}     
\label{bec:expan}
\lefteqn{\int_0^\infty \mathrm{d} q q^2 \ln \left(1 - {\rm e}^{-\beta E(q)}
\right) \stackrel{\mathrm{Taylor}}{\longrightarrow} } \nonumber \\ &&
\int_0^{\infty} \mathrm{d} q \left[ q^2 \, \ln\left(1 - {\rm
e}^{-q^2}\right) + \frac{q^2}{{\rm e}^{q^2} - 1} \bigl(\alpha + 4 y \bigr) -
\frac{1}{2} \frac{q^2 {\rm e}^{q^2}}{\left({\rm e}^{q^2} - 1\right)^2}
\bigl(\alpha + 4 y \bigr)^2 - \frac{1}{2} \frac{1}{{\rm e}^{q^2} - 1} 4 y^2
+ \mathcal{O}\left(\beta^3 \right) \right] \nonumber \\ && \quad = -
\tfrac{1}{2} \Gamma(\tfrac{3}{2}) \left[ \zeta (\tfrac{5}{2}) - \zeta
(\tfrac{3}{2}) \bigl(\alpha + 4 y \bigr) + \tfrac{1}{2} \zeta (\tfrac{1}{2})
\bigl(\alpha + 4 y \bigr)^2 + \zeta (\tfrac{1}{2}) 4 y^2 +
\mathcal{O}\left(\beta^3 \right) \right].
\end{eqnarray}
Contrary to the previous cases, we now have two dimensionless expansion
parameters $\alpha$ and $y$, both being proportional to $\beta$.  

At two loop, we consider the class of bubble diagrams (see Fig.~4 of
Ref.~\cite{effbos}) giving the following contributions to the
thermodynamic potential due to the gapped thermal modes:
\begin{equation} 
\frac{\Omega_2(\Phi)}{V} = - \frac{g}{4} (3 I_1^2 + 3 I_2^2 + 2 I_1 I_2),
\end{equation} 
where $I_1$ is the one-loop integral
\begin{equation} 
I_1 = \int \frac{\mathrm{d}^3 k}{(2\pi)^3} \frac{1}{E(k)} \frac{\hbar^2
k^2/2m - \mu + 2 g |\Phi|^2}{{\rm e}^{\beta E(k)}-1},
\end{equation} 
and a similar equation for $I_2$ with $2 g$ replaced with $6 g$ in the
numerator.  Expanding the integrals in a high-temperature expansion, we
obtain as two-loop contributions
\begin{equation}
\frac{\Omega_2(\Phi)}{V} = \frac{2 g}{\lambda^6} \left[\zeta^2(\tfrac{3}{2})
- 2 \zeta(\tfrac{1}{2}) \zeta(\tfrac{3}{2}) (\alpha + 4 y) \right] ,
\end{equation} 
the leading term of which was considered in Ref.~\cite{effbos}, but not the
rest.

Adding the contributions together, we arrive at
\begin{equation} \label{bec:lanform}
\frac{\Omega_0(\Phi) + \Omega_1(\Phi) + \Omega_2(\Phi)}{V} = c_0 - c_2
|\Phi|^2 + c_4 |\Phi|^4,
\end{equation} 
with 
\begin{equation}
\label{c0}
c_0 = - \frac{1}{\beta \lambda^3} \left\{ \zeta (\tfrac{5}{2}) - 2
\zeta^2(\tfrac{3}{2}) \delta - \zeta (\tfrac{3}{2}) \alpha \left[1- 4
\zeta(\tfrac{1}{2})\delta \right] + \tfrac{1}{2} \zeta (\tfrac{1}{2})
\alpha^2 \right\},
\end{equation}
a $\Phi$-independent term, with expansion parameter
\begin{equation} 
\label{delta}
\delta = g \beta/\lambda^3 = a/\lambda.
\end{equation}   
For $\delta \to 0$, Eq.~(\ref{c0}) is analogous to the first contributions
due to the gapped thermal modes of a free Bose gas given by Eq.~(\ref{Fint})
with Eq.~(\ref{highT}), albeit with $\alpha>0$ now.  Furthermore,
\begin{subequations}
\label{bec:c's}
\begin{eqnarray}
\label{bec:c1} 
-\beta c_2 &=& \left[\alpha + 4 \zeta (\tfrac{3}{2}) \delta \right] \left[1-
4 \zeta(\tfrac{1}{2}) \delta \right] , \\
\label{bec:c2} 
c_4 &=& g \left[1 - 12 \zeta (\tfrac{1}{2}) \delta  \right].
\end{eqnarray} 
\end{subequations}

It is important to note that one cannot naively set $\Phi$ to zero in
Eq.~(\ref{bec:lanform}) [with $\Omega_2(\Phi)$ omitted], as was done by Baym
\textit{et al.}~\cite{Baymetal_rev} to argue that the zeta function
regularization of infrared divergences used to obtain that result is flawed,
and hope to retrieve the correct critical behavior of the compressibility,
say, of a noninteracting Bose gas.  As explained in Sec.~(\ref{sec:highT}),
universal properties are determined exclusively by the fluctuations in the
zero-frequency mode.  These fluctuations are, however, not considered in
this section.  We are interested here in the condensation temperature--a
nonuniversal quantity.  As argued in the Introduction, nonuniversal
quantities in principle also depend on the gapped thermal modes.  It is the
effect of fluctuations in these dynamic modes that is summarized by
Eq.~(\ref{bec:lanform}).  More specifically, the coefficients given in
Eqs.~(\ref{bec:c's}) express the dressing of the parameters $\mu$ and $g$ by
these thermal fluctuations.  The one-loop corrections, first
calculated in Ref.~\cite{effbos}, have recently been rederived in
Ref.~\cite{AMT}, apparently in ignorance of the existing literature on the
subject. 

The condensation temperature is determined by the vanishing of the
coefficient $c_2$, yielding
\begin{equation}
\label{alphac} 
\alpha_\textrm{c} = -4 \zeta(\tfrac{3}{2}) \delta,
\end{equation} 
which gives $T_{\rm c}$ in terms of $\mu$.  Experimentally, it is more
realistic to consider the system at fixed particle number density and
therefore take $n$ rather than $\mu$ as independent variable.  The particle
number density is easily obtained from Eq.~(\ref{bec:lanform}) at
$T=T_\mathrm{c}$ where $\Phi$ vanishes:
\begin{equation}
\label{nmu}
n= - \frac{1}{V} \frac{\partial \Omega}{\partial \mu} \Biggr|_{T=T_{\rm c}}
= \frac{1}{\lambda_\mathrm{c}^3} \left\{\zeta(\tfrac{3}{2})\left[1- 4
\zeta(\tfrac{1}{2})\delta \right] - \zeta(\tfrac{1}{2}) \alpha_\mathrm{c}
\right\}.
\end{equation}
Together with Eq.~(\ref{alphac}) this reduces to the free Bose gas expression:
\begin{equation}
\label{nc}
n  = \zeta(\tfrac{3}{2})/\lambda_\mathrm{c}^3,
\end{equation}
so that to this order the dynamic modes do not lead to a shift in the
condensation temperature.  This conclusion differs from our previous one in
Ref.~\cite{effbos}, where the two-loop contributions were not properly
included \cite{Arnold}.  The additional two-loop contributions cancel the
shift due to the gapped thermal modes found in Ref.~\cite{effbos} at one
loop.

With the expression (\ref{nc}) for $n$, the chemical potential at
$T_\mathrm{c}$ can be written as
\begin{equation} 
\label{muc}
\mu_\mathrm{c} = 4 g n = \frac{8 \pi \hbar^2 a}{m} n,
\end{equation} 
which is nothing but the Hugenholtz-Pines relation \cite{HP} between the
chemical potential and the self-energy at the condensation temperature where
$\Phi$ disappears.  It is identical to the zero-temperature relation
(\ref{mu0}) apart from a subtle factor of 2 that can be understood by
examining the Hugenholtz-Pines relation, which is an exact result, at both
temperatures.  It follows from Eq.~(\ref{muc}) that the chemical potential
remains positive at the condensation temperature of a weakly interacting
Bose gas.  This is in contrast to a free Bose gas (obtained by letting $a
\to 0$), where the chemical potential vanishes when the condensation
temperature is approached from above.  Whereas the chemical potential in a
free Bose gas remains zero all the way down to zero temperature, in a weakly
interacting Bose gas it decreases from $4gn$ at $T_\mathrm{c}$ to $2gn$ at
zero temperature.

To justify the high-temperature expansion used, note that to lowest
nontrivial order, the coefficient $c_2$ is given by $-\beta c_2 = \alpha + 4
\zeta (\tfrac{3}{2}) \delta$.  The condition $c_2=0$ then gives the
condensation temperature in terms of $\mu$ to lowest order as \cite{effbos}
\begin{equation} 
\label{consi}
k_\mathrm{B} T_\mathrm{0} = \frac{2 \pi \hbar^2}{m} \left(\frac{\mu}{4 g
\zeta(\tfrac{3}{2})} \right)^{2/3},
\end{equation} 
which is a one-loop result.  Since this temperature is large for $g$
small, the high-temperature expansion is consistent with the
weak-coupling assumption of perturbation theory.  Using the
Hugenholtz-Pines relation (which is also satisfied at this lowest
nontrivial order) to replace $\mu$ with $n$, Eq.~(\ref{consi}) takes the
standard form (\ref{T0d}) for a noninteracting Bose gas, independent of
$g$.  As for a free Bose gas, this result is determined entirely by the
nonzero frequency modes.

Because $\alpha_c \propto \delta \propto g$, the coefficients (\ref{bec:c1})
and (\ref{bec:c2}) of the effective theory are close to $T_\mathrm{c}$
calculated up to the order $g^2$.  In the first coefficient (\ref{c0}),
the term $\propto \delta^2$ is missing.  This term, given by the leading
contribution in the high-temperature expansion of the third-order bubble
diagram, is independent of the chemical potential and therefore irrelevant
for our purposes.

\subsection{BEC in two dimensions}
\label{sec:generald}
For arbitrary dimension $2<d<4$, Eqs.~(\ref{alphac}) and (\ref{nc})
generalize to:
\begin{equation}
\label{ncd}
\alpha_\textrm{c} = -4 \zeta(d/2) \delta , \;\;\; n =
\zeta(d/2)/\lambda_\mathrm{c}^d,
\end{equation}
where now $\delta = g \beta/\lambda^d$.  The limiting case, $d=2$, is
special because $\zeta(d/2)$ diverges when $d \to 2$.  To
investigate this limit, we dimensional regularize the last equation by
considering the problem in $d = 2 + \epsilon$, where
\begin{equation} 
\label{tiny}
n = \frac{2}{\lambda_\mathrm{c}^2} \frac{1}{\epsilon} .
\end{equation} 
As in Eq.~(\ref{connect}), the pole in dimensional regularization can be
connected with the logarithm appearing in the regularization with a momentum
cutoff, provided the ultraviolet momentum cutoff and the relevant infrared
scale are identified.  Here, they are given by the inverse range of the
potential $1/a$ and the square root of the chemical potential, respectively
\cite{FisherHohenberg}, i.e.,
\begin{equation} 
\frac{1}{\epsilon } \rightarrow \ln \left(\frac{\hbar}{a \sqrt{m \mu}}
\right) = - \frac{1}{2} \ln \left( \frac{8 \pi n a^2}{\ln(1/8 \pi n
a^2)} \right) \approx \frac{1}{2} \ln[ \ln (1/8 \pi n a^2)]
\end{equation} 
for $\ln[ \ln (1/8 \pi n a^2)]>>1$.  In deriving this use is made of the
two-dimensional relation between the chemical potential and the particle
number density at the condensation temperature, 
\begin{equation} 
\label{muln}
\mu_\mathrm{c} = \frac{8 \pi \hbar^2 n}{m \ln(1/8 \pi n a^2)}.
\end{equation} 
With this correspondence, Eq.~(\ref{tiny}) leads to the well-known
expression for the critical temperature \cite{Popov,FisherHohenberg}
\begin{equation} 
\label{T2}
k_\mathrm{B} T_\mathrm{c} = \frac{2 \pi \hbar^2 n}{m \ln[ \ln (1/8 \pi n
a^2)]}.
\end{equation} 

Together with the relation $\mu_\mathrm{c} = 4 g n$, Eq.~(\ref{muln}) gives
for the coupling constant
\begin{equation}
\label{g} 
g = \frac{2 \pi \hbar^2}{m \ln (\hbar^2/m \mu a^2)}.
\end{equation} 
To understand this result, recall that the time dimension counts double
compared to the space dimensions in a nonrelativistic quantum theory.  The
quantum critical behavior of the theory was first investigated by Uzunov
\cite{Uzunov}, who showed that $d=2$ (implying a total of four effective
spacetime dimensions) corresponds to the upper critical dimension, below
which a nontrivial infrared stable fixed point exists, with $g>0$.  For
$d>2$, the theory has a trivial infrared stable fixed point at $g=0$.  The
logarithm in Eq.~(\ref{g}) arises because $d=2$ separates these two
different cases, with the chemical potential providing an infrared cutoff,
so that a finite value for $g$ and ultimately for $T_\mathrm{c}$ obtains.
In other words, the finite value for the condensation temperature in two
dimensions hinges on the presence of quantum fluctuations.

As an aside, we note that at zero temperature, the relation between the
chemical potential and particle number density differs again a subtle
factor of 2 from the relation $\mu_\mathrm{c} = 4 g n$ we used to derive
the critical temperature (\ref{T2}).  As in $d=3$, this factor is needed
for the Hugenholtz-Pines relation to be satisfied at both temperatures.
It implies that the analog of Eq.~(\ref{muln}) at zero temperature also
differs a factor of 2, which is indeed what is found by Fisher and
Hohenberg \cite{FisherHohenberg}.

\subsection{Discussion}
\label{sec:comp}
Path-integral Monte Carlo simulations by Gr\"uter, Ceperly, and Lalo\"e
\cite{GCL} of the microscopic model describing a weakly interacting Bose gas
restricted to relatively few (up to 216) particles in three dimensions found
an increase in the condensation temperature linear in the scattering length,
\begin{equation} 
\label{shift}
\frac{T_{\rm c} - T_0 }{T_0} = C a n^{1/3},
\end{equation}
with $C = 0.34 \pm 0.06$.  Since the dynamic modes do not contribute to
the shift linear in $a$, the coefficient $C$ can also be calculated in
the static, classical theory.  Recent Monte-Carlo simulations \cite{MC}
of that static theory gave larger values: $C = 1.32 \pm 0.02$ and $1.29
\pm 0.05$, respectively.  An intuitive understanding of the increase in
$T_\mathrm{c}$ was given in Ref.~\cite{GCL}, where it was pointed out
that a moderate repulsive interaction suppresses density fluctuations.
This results in a more homogeneous system and facilitates the formation
of large exchange rings necessary for BEC, which then takes place
already at a higher temperature than in a free Bose gas.  When the
repulsive interaction increases further, the exchange is obstructed
because it becomes more difficult for the particles to move.  This leads
to a lowering of the critical temperature as seen, for example, in
liquid $^4$He.  A free gas with $^4$He parameters at vapor pressure
would have a condensation temperature of about 3 K, whereas liquid
$^4$He becomes superfluid at the lower temperature of 2.17 K.

The full thermodynamic potential of an interacting Bose system can close
to $T_\mathrm{c}$ be written in terms of the exchange rings by
slightly generalizing Eq.~(\ref{loopsf}) for a free Bose gas
\cite{percolation}:
\begin{equation} 
\frac{\Omega}{V} \propto \sum_{w=1}^\infty w^{- \tau}
\mathrm{e}^{-(\alpha-\alpha_\mathrm{c}) w },
\end{equation} 
where now the exponent $\tau$ has the general form
\begin{equation} 
\tau = \frac{d}{D}+1,
\end{equation}  
with $D$ the fractal dimension of the worldlines, which is $D=2$ for a
free Bose gas.  The worldline tension determined by
$\alpha-\alpha_\mathrm{c}$ vanishes when the condensation temperature is
reached from above as
\begin{equation} 
\alpha-\alpha_\mathrm{c} \propto (T-T_\mathrm{c})^{1/\sigma},
\end{equation} 
where $\sigma$ is a second exponent, which for a free Bose gas takes the
value $\sigma= d/2-1$ [see above Eq.~(\ref{compr})].  The two exponents
$\sigma$ and $\tau$ determine all the critical exponents characterizing
the phase transition through scaling relations \cite{percolation}.  BEC
in a weakly interacting Bose gas is in the same universality class as
the lambda transition of liquid $^4$He, so that the critical exponents
describing BEC are accurately known.  In going from a free Bose gas to
strongly interacting $^4$He, the value of the fractal dimension $D$ of
the worldlines changes only slightly from $D=2$ to $D \approx 1.96$,
with most of the interaction effects entering $\sigma$, which changes
from $\sigma = 1/2$ to $\sigma \approx 0.76$ \cite{percolation}.  The
value of the fractal dimension follows from the relation
\cite{percolation} $D = 2 -\eta$, with $\eta$ the Fisher exponent,
determining the anomalous dimension of the order parameter $\Phi$.

For an ideal Bose gas, the value $D=2$ derives from the quadratic form
of the energy spectrum $E(k) = \hbar^2 k^2/2m$.  A different value is
obtained by modifying the free spectrum to $E(k) \sim k^D$, with $D \neq
2$.  It was noticed by Gunton and Buckingham \cite{GuntonBuckingham}
that an ideal gas with $D=3/2$, so that $E(k) \sim k^{3/2}$, produces a
value of the specific-heat critical exponent $\alpha_\mathrm{ce}=0$
close to the experimental value $\alpha_\mathrm{ce} \approx -0.01$ for
$^4$He \cite{cv4}.  However, this is a somewhat fortuitous and at the
same time deceptive coincidence as other exponents come out incorrectly,
and the actual value of the fractal dimension is close to 2, $D \approx
1.96$ \cite{percolation}.  Describing the critical behavior of an
interacting Bose gas, using quasiparticles with the spectrum
\begin{equation} 
E(k) \sim k^{3/2},
\end{equation}  
as was done by Baym \textit{et al.} \cite{Baymetal}, can therefore not
be justified.  

In addition to numerical studies of the static, classical theory, the
coefficient $C$ has also been estimated by analytical studies of that theory
(for a summary, see Ref.~\cite{Baymetal_rev}), such as the $1/N$ expansion
\cite{BBZ} and variational perturbation theory \cite{Kleinert_tc}.

Although the dynamic modes do not, due to cancellations, contribute to
the shift in the condensation temperature linear in $a$, they do
determine the coefficient of the next term $\sim a^2 n^{2/3} \ln(a
n^{1/3})$ in that shift \cite{AMT}.

\section{Conclusions}
\label{sec:concl}
In this paper, the effect of fluctuations in the nonzero frequency modes
of a weakly interacting Bose gas on the condensation temperature was
studied in detail.  The simple algorithm used to perturbatively
calculate the effect in a high-temperature expansion in conjunction with
zeta function regularization of infrared divergences was demonstrated to
be reliable by showing that a host of known results are recovered.  The
presence of temperature-induced energy gaps for these dynamic modes were
argued to allow for a perturbative approach.  It was shown that the
shift in the condensation temperature of the form (\ref{shift}) with
\begin{equation} 
\label{per}
C = - \frac{8}{3} \frac{\zeta(\frac{1}{2})}{\zeta^{1/3}(\frac{3}{2})}
\approx 2.83,
\end{equation}
we had obtained earlier at one loop \cite{effbos} is canceled by
two-loop contributions \cite{Arnold}.

\begin{acknowledgments} 
The author acknowledges helpful correspondence with and critical comments of
P. Arnold, M. Haque and H. Kleinert.
\end{acknowledgments}

}


\begin{thebibliography}{99}
\bibitem{Arnold} This cancellation was pointed out to the author by
P. Arnold, see also Ref.~\cite{AMT}.
\end{thebibliography}
\end{document}